# COMPARING THE IMPACT OF MOBILE NODES ARRIVAL PATTERNS IN MANETS USING POISSON AND PARETO MODELS

John Tengviel[1], and K. Diawuo[2]

[1]Department of Computer Science, Sunyani Polytechnic, Sunyani, Ghana
john2001gh@yahoo.com
[2]Department of Computer Engineering, KNUST, Kumasi, Ghana
kdiawuo@yahoo.com

## ABSTRACT

*Mobile Ad hoc Networks (MANETs) are dynamic networks populated by mobile stations, or mobile nodes (MNs). Mobility model is a hot topic in many areas, for example, protocol evaluation, network performance analysis and so on. How to simulate MNs mobility is the problem we should consider if we want to build an accurate mobility model. When new nodes can join and other nodes can leave the network and therefore the topology is dynamic. Specifically, MANETs consist of a collection of nodes randomly placed in a line (not necessarily straight). MANETs do appear in many real-world network applications such as a vehicular MANETs built along a highway in a city environment or people in a particular location. MNs in MANETs are usually laptops, PDAs or mobile phones.*

*This paper presents comparative results that have been carried out via Matlab software simulation. The study investigates the impact of mobility predictive models on mobile nodes' parameters such as, the arrival rate and the size of mobile nodes in a given area using Pareto and Poisson distributions. The results have indicated that mobile nodes' arrival rates may have influence on MNs population (as a larger number) in a location. The Pareto distribution is more reflective of the modeling mobility for MANETs than the Poisson distribution.*

## 1. INTRODUCTION

Mobile Ad-hoc Networks (*MANETs)* is a collection of wireless mobile nodes configured to communicate amongst each other without the aid of an existing infrastructure. MANETS are *Multi-Hop* wireless networks since one node may not be indirect communication range of other node. In such cases the data from the original sender has to travel a number of hops (hop is one communication link) in order to reach the destination. The intermediate nodes act as routers and forward the data packets till the destination is reached [1].

Recently, with the deployment of all kinds of wireless devices, wireless communication is becoming more important. In this research area, Ad-Hoc network is a hot topic which has attracted much of research attentions. A wireless ad hoc network is a decentralized wireless network. The network is ad hoc because it does not rely on a preexisting infrastructure, such as routers in wired networks or access points in managed (infrastructure) wireless networks. Instead, each node participates in routing by forwarding data for other nodes, and so the determination of which nodes forward data is made dynamically based on the network connectivity [2]. There are different kinds of routing protocol defined by how messages are sent from the source node to the destination node.

Based on this, it's reasonable to consider node mobility as an essential topic of ad-hoc network. With an accurate mobility model which represents nodes movement, designers can evaluate performance of protocols, predict user distribution, plan network resources allocation and so on. It can also be used in healthcare or traffic control area rescue mission, and so on.

Ad hoc networks are viewed to be suitable for all situations in which a temporary communication is desired. The technology was initially developed keeping in mind the military applications [3] such as battle field in an unknown territory where an infrastructure network is almost impossible to have or maintain. In such situations, the ad hoc networks having self-organizing [4] capability can be effectively used where other technologies either fail or cannot be effectively deployed. The entire network is mobile, and the individual terminals are allowed to move freely. Since, the nodes are mobile; the network topology is thus dynamic. This leads to frequent and unpredictable connectivity changes. In this dynamic topology, some pairs of terminals may not be able to communicate directly with each other and have to rely on some other terminals so that the messages are been delivered to their destinations. Such networks are often referred to as multi-hops or store-and-forward networks [5].

This paper presents a study on mobile nodes arrival patterns in MANETs using Poisson and Pareto models. Though not very realistic from a practical point of view, a model based on the exponential distribution can be of great importance to provide an insight into the mobile nodes arrival pattern. The section 2 illustrates a brief review on MANETs studies. The section 3 introduces the Poisson and Pareto distribution models. The simulation procedures and considered parameters are presented in section 4. The obtained results are objects in section 5 and the section 6 closes the paper to further research works.

## 2. RELATED WORKS

Currently there are two types [6, 7] of mobility models used in simulation of networks. These are traces and synthetic models. Traces are those mobility patterns that are observed in real-life systems. Traces provide accurate information, especially when they involve a large number of mobile nodes (MNs) and appropriate long observation period. On the other hand, synthetic models attempt to realistically represent the behaviour of MNs without the use of traces. They are divided into two categories, entity mobility models and group mobility models [1, 8, 9]. The entity mobility models randomise the movements of each individual node and represent MNs whose movements are independent of each other. However, the group mobility models are a set of groups' nodes that stay close to each other and then randomise the movements of the group and represent MNs whose movements are dependent on each other. The node positions may also vary randomly around the group reference point. In [10], the mobility study in ad hoc has been approximated to pedestrian in the street, willing to exchange content (multimedia files, mp3, etc.) with their handset whilst walking at a relative low speed. Some researchers have proposed basic mobility models such as Random Walk, Random Waypoint, [3, 4], etc. for performance comparison of various routing protocols. The concern with these basic designed models is that they represent a specific scenarios not often found in real lives. Hence their use in ad hoc network studies is very limited. Random Walk or Random Waypoint model though simple and elegant, produce random source of entry into a location with scattered pattern around the simulation area, sudden stops and sharp turns. In real-life, this may not really be the case.

## 3. Models of Study

### 3.1. Poisson Arrival Distribution (Number of Nodes)

When arrivals occur at random, the information of interest is the probability of *n* arrivals in a given time period, where *n* = 0, 1, 2, . ……n-1

Let λ be a constant representing the average rate of arrival of nodes and consider a small time interval Δ**t**, with Δ**t** →**0**. The assumptions for this process are as follows:

- The probability of one arrival in an interval of Δt seconds, say **(t, t+Δt)** is λΔ**t**, independent of arrivals in any time interval not overlapping **(t, t+Δt).**
- The probability of no arrivals in Δ**t** seconds is **1-**λΔt, under such conditions, it can be shown that the probability of exactly **n** nodes arriving during an interval of length of **t** is given by the Poisson distribution law [11] in equation 1:

$$\mathbf{P(n)} = \frac{(\lambda t)^n e^{-\lambda t}}{n!}, \text{ where } n \geq 0,\ t > 0 \ . \qquad (1)$$

The assumption of Poisson MN arrivals also implies a distribution of the time intervals between the arrivals of successive MN in a location.

### 3.2. Pareto Distribution

The Pareto distributions [12-14] are characterized by two parameters: $\alpha$ and $\beta$. Parameter $\alpha$ is called shape parameter that determines heavy-tailed characteristics and $\beta = 1$ is called cutoff or the location parameter that determines the average of inter-arrival time.

The node arrival times of the Pareto distribution are independent and identically distributed, which means that each arrival time has the same probability distribution as the other arrival times and all are mutually independent. The two main parameters of the Pareto process are the shape $\alpha$ and the scale parameter (x).

For one parameter Pareto ($\alpha$ shape only), the distribution function can be written as equation 2:

$$F(X) = 1 - \left(\frac{1}{1+X}\right)^{\alpha}, X \geq 0 \qquad (2)$$

The pdf is given as in equation 3:

$$f(X) = \frac{\alpha}{(1+X)^{\alpha+1}} \qquad (3)$$

and for the two – parameter Pareto distribution function defined over the real numbers can be written as in (4):

$$\begin{cases} F(X) = 1 - \left(\frac{1}{\alpha+X}\right)^{\beta} \\ \quad , \\ X \geq 0; \quad \alpha, \beta > 0 \end{cases} \qquad (4)$$

Its pdf is given as in equation 5:

$$f(X) = \frac{\alpha}{\beta} * \left(\frac{\beta}{X}\right)^{\alpha} \qquad (5)$$

# 4. METHODOLOGY

## 4.1. Varying of α in Pareto Arrival Distribution

We assume the arrival distribution on the MNs population by using Pareto distributions.

**Table 1: Varying α parameter values**

| Scenario | 1 | 2 | 3 | 4 | 5 |
|---|---|---|---|---|---|
| α (B) | 0.3 | 0.<br>4 | 0<br>.5 | 0.<br>8 | 0.9 |

For the simulations purposes, the varying α values are been considered. Heavy-tail is been modeled by a Pareto distribution and the main principle can be attributed to the principle of number of nodes. We have performed the simulations for a wide range of parameter values as in Table 1 for both one-parameter and two-parameter Pareto models.

## 4.2. VARYING OF ARRIVAL RATES FOR NODE DISTRIBUTION

The arrival pattern of mobile nodes has an impact on the performance of the network. In this scope, we have decided to analysis the effect of arrival distribution on the MNs population in a given area by using Poisson distribution as in equation 1. In most real-world MANETs, the node population in an area of interest varies with time. In this simulation, it is therefore necessary to investigate the impact of arrivals of MNs on the MANETs mobility.

The simulation area does not change as the arrival rate changes. The different values of arrival rates being considered in this study are shown in Table 2.

**Table 2: Varying Arrival Rates**

| Scenario | 1 | 2 | 3 | 4 | 5 |
|---|---|---|---|---|---|
| Arrival rates | 0.<br>3 | 0.<br>4 | 0<br>.5 | 0.<br>8 | 0.9 |

During the simulation, nodes were allowed to enter the location from a common source (0 degrees) but not from different sources. The number of MNs that entered the location was assumed to be Poisson distributed with varying arrival rates.

# 5. RESULTS AND DISCUSSION

## 5.1. Comparative Study using Pareto Arrival Pattern

In this section, the effect of arrival rates on MNs distribution and population in a defined location is analyzed as shown in Figure 1. It was observed that the various arrival rates increased the number of MNs also increased but to a certain limit. It is therefore the indication that every location has a limit or capacity of MNs it can contain.

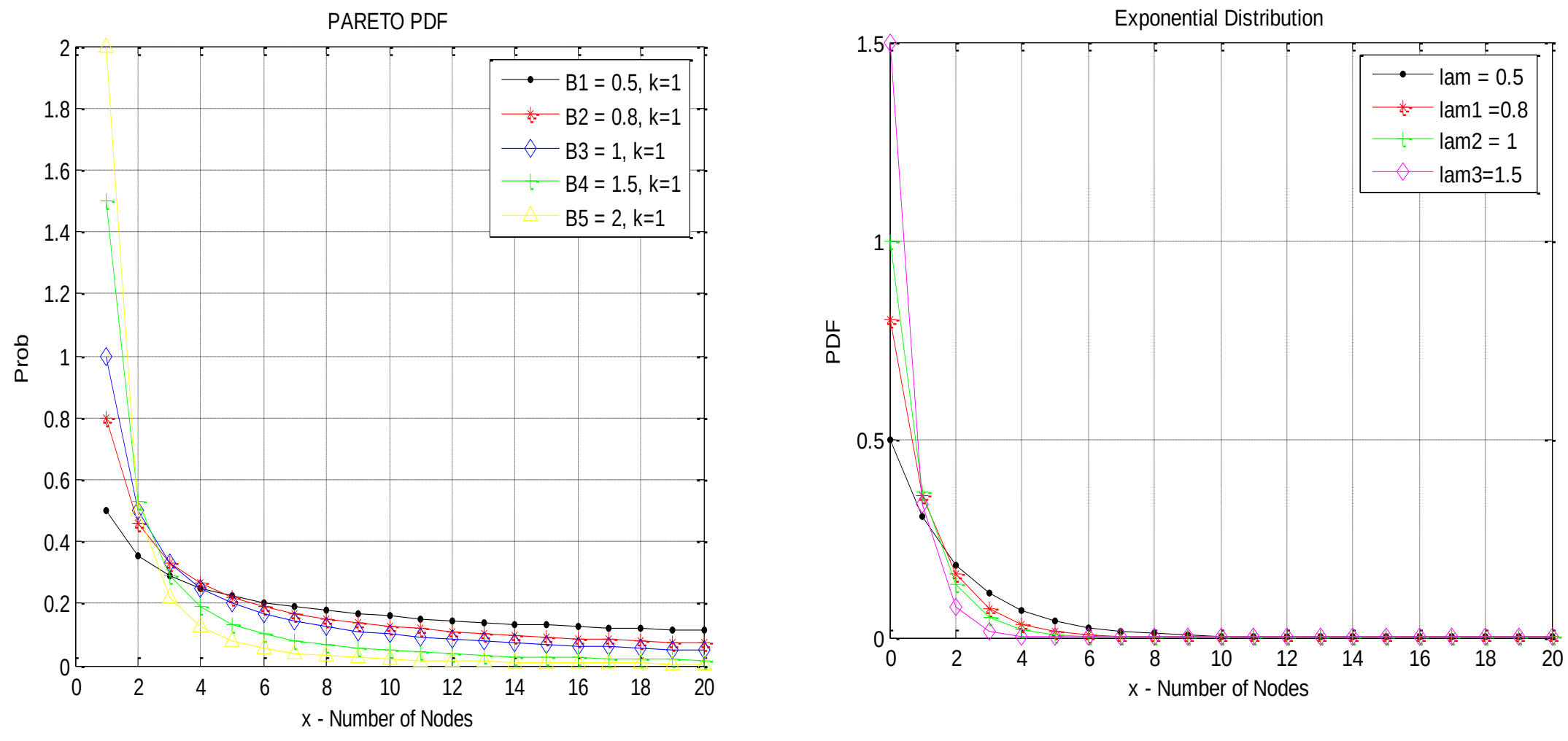

**Figure 1: Single Parameter for Varying Values for B, and Exponential for Twenty Nodes**

Figure 1 may indicate that the exponential distribution was higher than the single parameter in the initial stages but as time progresses the exponential decreases fast to zero. The single parameter Pareto overtakes the exponential as the number of nodes increases and indication that the single parameter performs better than exponential distribution.

The Pareto distribution may show tail that decays much more slowly than the exponential distribution. The alpha is the shape parameter which determines the characteristics "decay" of the distribution (tail index) and A is the location parameter which defines the minimum value of x (number of nodes).

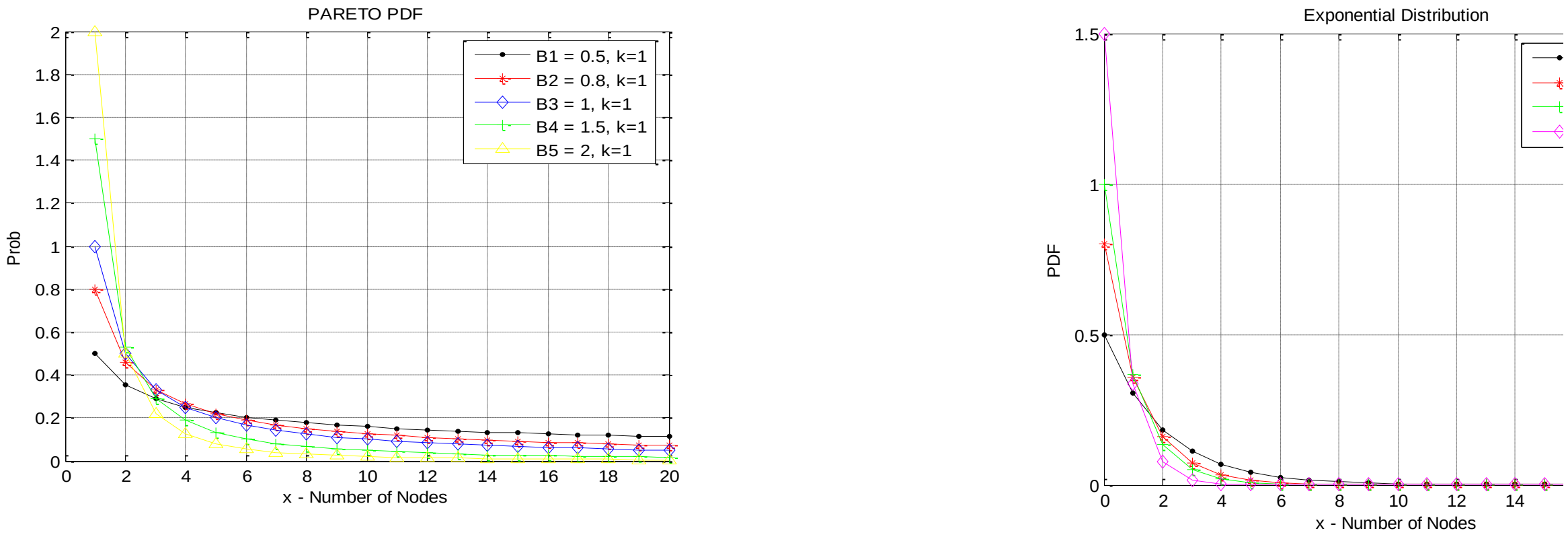

**Figure 2: Two Parameter Pareto for Varying B Values an**

In Figure 2 the comparison between the two-parameter Pareto and exponential distributions is illustrated. It is obvious that the two-parameter Pareto outweighs the exponential distribution as the number of MNs increases. The exponential distributions decays very fast and finally get to the a-axis unlike the two-parameter Pareto distribution where some of the arrival rates distribution has not decay to zero.

However the two-parameter Pareto performed well than the one-parameter Pareto, since some of the arrival of the two-parameter did not decay to zero. The long-tailed nature of the two-parameter Pareto helped to clear out any congestion in a location when the arrival rate was small and the reverse was also true.

## 5.2 Effect of Varying Arrival Rates

In Figure 3, the effect of varying nodes' arrival rate is computed using Poisson model. Nodes may arrive at a location either in some regular pattern or in a totally random fashion. The arrival rates have shown to impact on the number of nodes in a particular location, although every location has a limited capacity. A high number of nodes typically translate into a higher average number of neighbours per node, which influences the route availability.

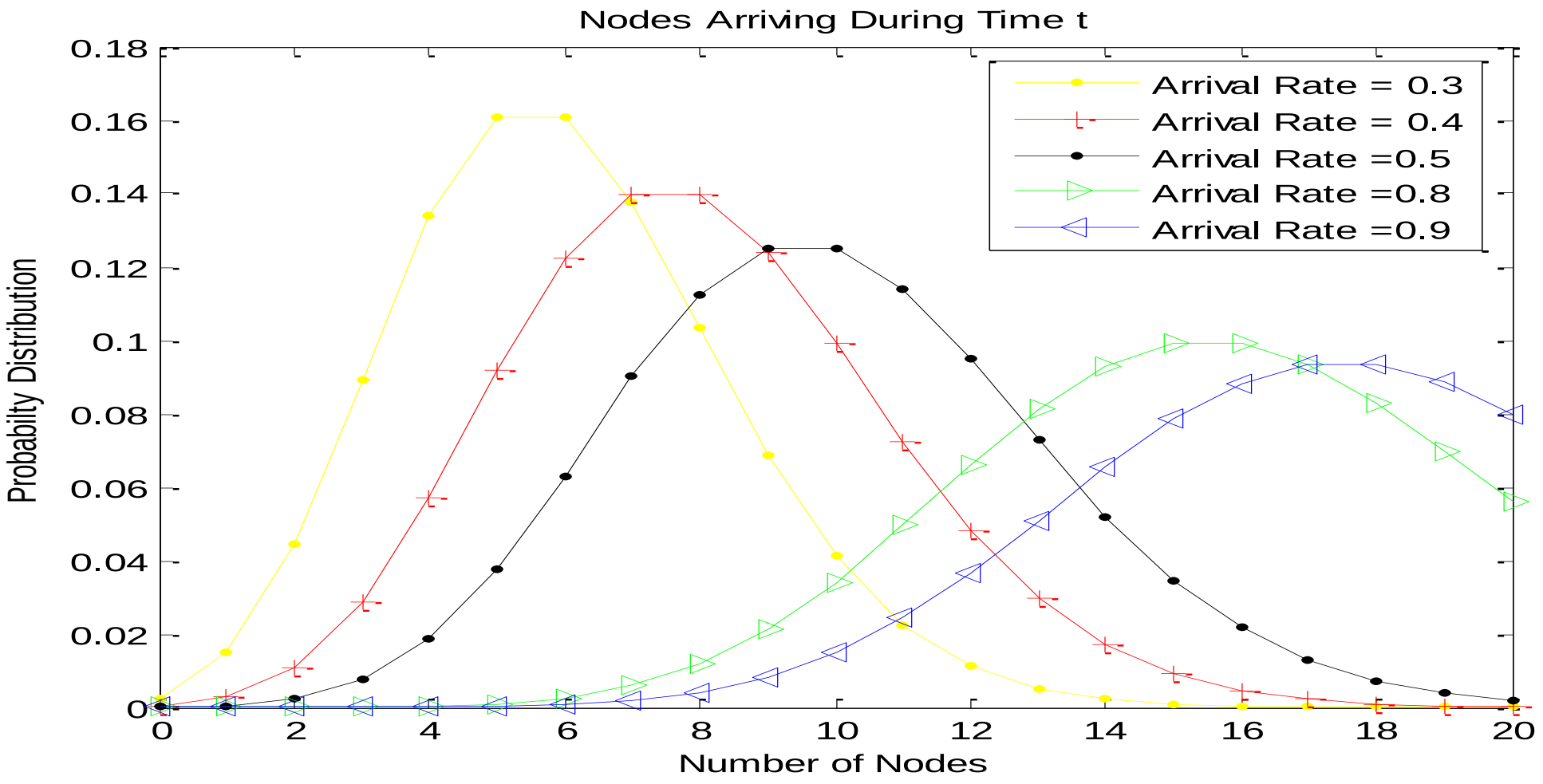

**Figure 3: For Twenty Number of Nodes for varying Arrival rates**

In reality, the total connection time of a node over a specific interval depends on the nodes encounter rate and the time in each encounter, both of which depend on the relative mobility of nodes.

Although a high node arrivals results in more node encounters, the network would eventually become congested. The impact of this relationship is that nodes can and will be tightly packed (i.e. High density) if their arrival rates is high (congestion), but if the arrivals is lower, the nodes must be farther apart (low density). For instance it is clear that there is some congestion for arrivals of MNs, since they have to follow some holding paths.

As the value of arrival rate increases, the shape of the distribution changes dramatically to a more symmetrical ("normal") form and the probability of a larger number of arrivals increases with increasing number of MNs. An interesting observation is that as the arrival rate increases, the properties of the Poisson distribution approach those of the normal distribution as in Figure 3.

The first arrival processes of nodes give higher contact probabilities at higher arriving rates. This is due to the nodes' contiguity one to another making mobility difficult. In practice, one may record the actual number of arrivals over a period and then compare the frequency of distribution of the observed number of arrival to the Poisson distribution to investigate its approximation of the arrival distribution.

# 6. Conclusion

The arrival patterns have shown some impact on the network population, as the arrival rate increases the MNs population also increases to a peak and then decays rapidly to the x-axis. It was realized that the Poisson distribution is not good for the arrival distribution; therefore the Pareto distribution was considered. It has come out clear that the Pareto distribution is good for the arrival distribution, especially the two-parameter Pareto distribution which performed better than the single Pareto and exponential distributions even though at the earlier stages the exponential performed than the single Pareto distribution with a faster decay.

It may subsequently be admitted that mobility in MANETs is a difficult work and actually. It is an interesting research area that has been growing in recent years. Its difficulty is mainly generated because of the continuous changes in the network topology with time. The topological changes have impact on mobility techniques developed for infrastructure-based networks thus may not be directly applied to mobile adhoc networks. We have investigated through simulation mobility prediction of MNs using the queueing model.

**Authors**

**John Tengviel**

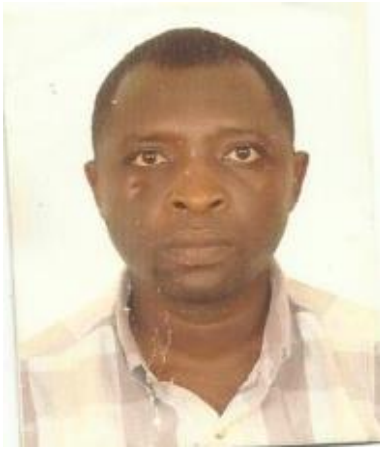

He is a holder of a BSc. Computer Science from Kwame Nkrumah University of Science and Technology (KNUST) in 2001 and MSc. Telecommunication Engineering from College of Engineering at the same university in 2012. He is currently a Lecturer with the Department of Computer Science at Sunyani Polytechnic. His research interests include Mobile Ad hoc Networks, Wireless Communication, Mobility Modeling, MANETs and Database System.

Nana (Dr.) Kwasi Diawuo is a senior lecturer of the Department of Computer Engineering at Kwame Nkrumah University of Science and Technology (KNUST), Kumasi, Ghana. He earned a BSc. (Electrical/ Electronic Engineering) from KNUST, M.Sc., Ph.D, and MGhIE. He is a member of the Institution of Electrical and Electronic Engineers (IEEE) and Computer Society (of IEEE).